\documentclass[aip,jcp,10 pt,reprint]{revtex4-1}
\usepackage[pdftex]{graphicx}
\usepackage{amsmath,amsfonts,amssymb,booktabs}
\usepackage{topcapt}
\usepackage{epstopdf}

\begin{document}





\title{Molecular internal dynamics studied by quantum path interferences in high order harmonic generation.} 








\author{Amelle Za\"ir}\email[]{azair@imperial.ac.uk}

\author{,
Thomas Siegel$^{1}$, Suren Sukiasyan $^{1}$, Francois Risoud $^{1}$, Leonardo Brugnera $^{1}$, Christopher Hutchison $^{1}$, Zsolt Diveki $^{2}$, Thierry Auguste $^{2}$, John W.G. Tisch$^{1}$, P. Sali\`eres $^{2}$, Misha Y. Ivanov $^{1}$, Jonathan P. Marangos}



\affiliation{$^{1}$Imperial College London, Department of Physics, Blackett Laboratory Laser Consortium, London, SW7 2AZ United Kingdom.\\
$^{2}$ Service des Photons, Atomes et Mol\'ecules, CEA-Saclay, 91191 Gif-sur-Yvette, France.
}






\begin{abstract}

We investigate how short and long electron trajectory contributions to high harmonic emission and their interferences give access to intra-molecular dynamics. In the case of unaligned molecules, we show experimental evidences that the long trajectory signature is more dependent upon the molecule than the short one, providing a high sensitivity to cation nuclear dynamics within 100's of as to few fs. Using theoretical approaches based on Strong Field Approximation and Time Dependent Sch\"odinger Equation, we examine how quantum path interferences encode electronic motion whilst molecules are aligned. We show that the interferences are dependent on channels superposition and upon which ionisation channel is involved. In particular, quantum path interferences encodes electronic migration signature while coupling between channels is allowed by the laser field.  
Hence, molecular quantum path interferences is a promising method for Attosecond Spectroscopy, allowing the resolution of ultra-fast charge migration in molecules after ionisation in a self-referenced manner.

\end{abstract}


\maketitle 

\section{Introduction}
In the intuitive description of the high-harmonic generation (HHG) mechanism using a linearly polarised laser field \cite{corkum_prl_1993, lewenstein_pra_1995} an initially bound wave packet is first freed in the continuum near the peak of the laser electric field (step 1: ionisation). During this step the system consists of a superposition of a continuum electronic wave packet (EWP) subject to the laser field and an EWP remaining in its ground state. The continuum EWP follows a trajectory where it is accelerated away and returned to the core driving by the oscillating laser field (step 2: propagation). At the returning time, this EWP can then interfere with the part that remained bound in the system and release the extra energy gained during the trajectory into a coherent harmonic photons emission (step 3: recombination). The harmonic photons emitted cover a large range of frequencies from UV to XUV and the corresponding spectra presents a specific extended comb of harmonic frequencies with a plateau followed by a cut-off region. This large spectral range allows experimentally the production of an isolated attosecond pulse \cite {hentschel_nature_2001, sansone_science_2007} or a train of XUV attosecond pulses\cite{paul_science_2001, mairesse_science_2003}. Classically, the instant at which the EWP is freed in the continuum, referred to as ionisation time, determines the trajectory the continuum EWP will follow during the propagation step.

 It is now well-established \cite{lewenstein_pra_1995, gaarde_np_2009, salieres_science_2001} that in the plateau region multiple families of trajectories can contribute to the same harmonic emission whilst in the cut-off (highest photon energies) one trajectory family contributes only, referred to as the 'cut-off trajectory'. To give a first estimation of these trajectories one can simply resolve the classical Newton equation of motion of a charge particle (electron) in a laser field over one single laser cycle as described in figure 1.
The three type of trajectories are presented in 1.a with a colour map representing the kinetic energy the electron acquired among a specific trajectory. This kinetic energy is then plotted as a function of ionisation and recombination time in 1.b. The maximum kinetic energy, which corresponds to the highest photon energy emitted, results from the cut-off trajectory. From this trajectory one can define within an optical cycle two families of trajectories, referred to as the 'short' (with duration smaller that the cut-off one) and the the 'long' (with duration longer that the cut-off one). 
The short trajectories are characterised by a kinetic energy that increases with the excursion time while the long trajectory has a kinetic energy that decreases as a function of the excursion time in the continuum. In both case this results in an attosecond chirp (positive for the short trajectory and negative for the long) that provides a time-frequency mapping. Therefore dynamical effects of the core (time domain) during HHG are encoded in the high harmonics spectra (frequency domain). 
For kinetic energy below the cut-off energy (maximum), there exists therefore a short and a long trajectory that both lead to the emission of the same photon energy. These are the two trajectories that can interfere leading to the method we propose here for studying Attosecond spectroscopy.
\begin{figure}[here]
	\includegraphics [width=3.2 in,height=1.8 in]{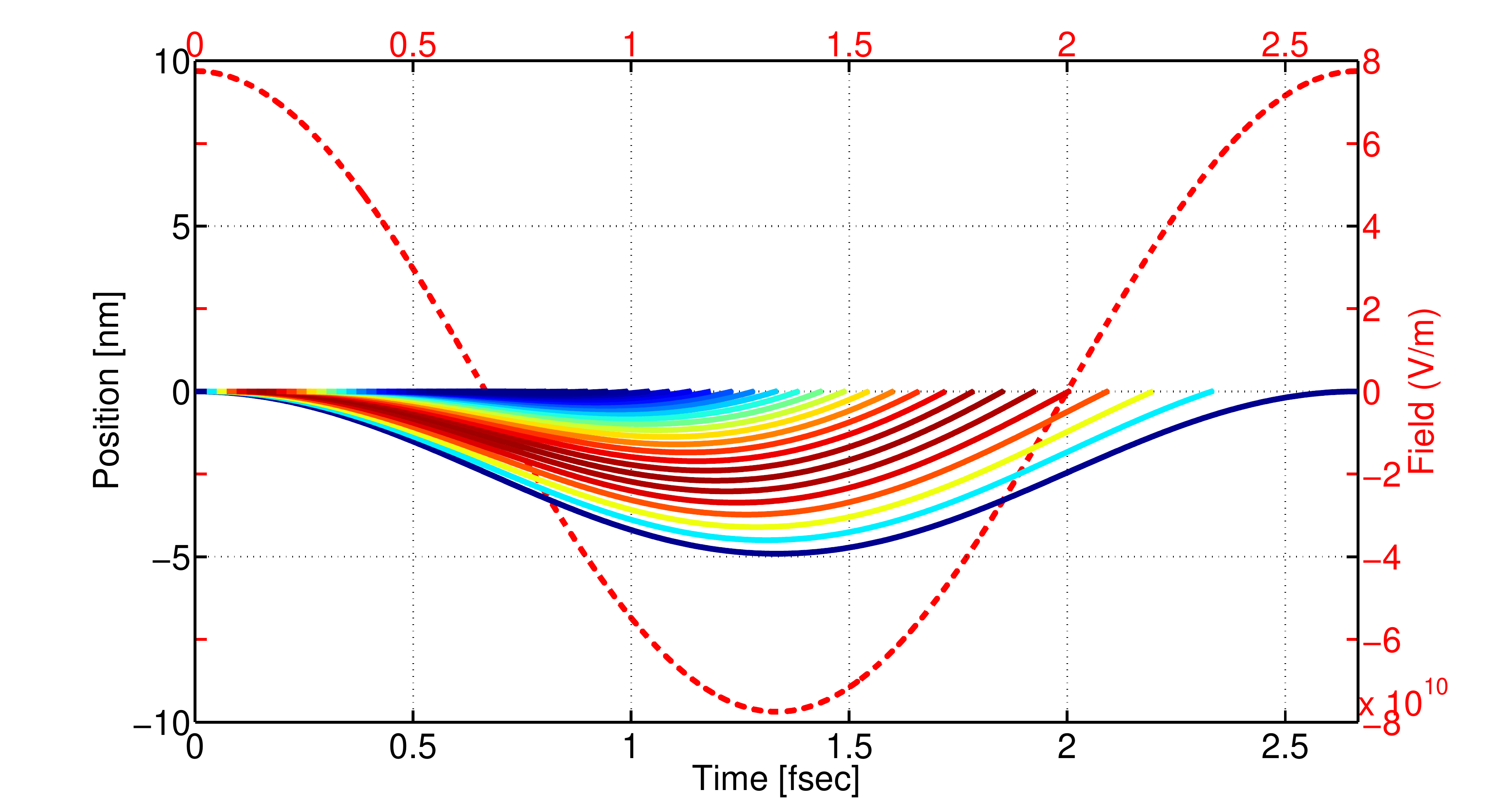}
	\includegraphics [width=3.2 in,height=1.8 in]{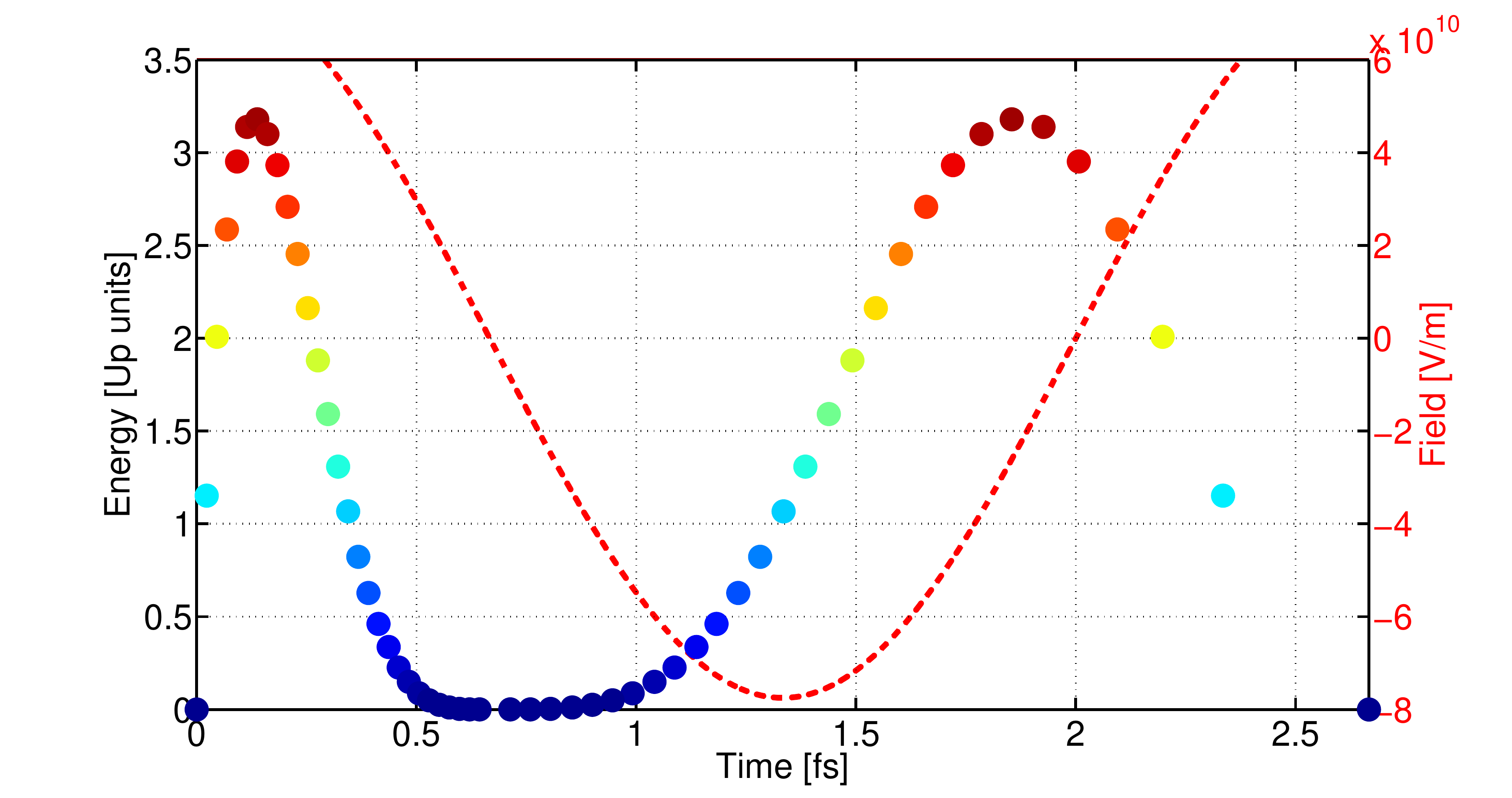}
\caption{Classical calculations: Left hand 1.a: trajectories as a function of the ionisation and returning time. Right hand 1.b: corresponding kinetic energy acquired by the electron freed in the laser field.
	Cut-off trajectory (red) corresponds to the highest kinetic energy leading to the so called cut-off law ($1.3I_{p}+3.2U_{p}$). At a given kinetic energy below the cut-off threshold, two trajectories referred to as 'short' and 'long', contribute to the same kinetic energy}
	\label{fig:fig1}
\end{figure}

In the quantum picture and under the strong field approximation these trajectories correspond to Feynman’s paths where the harmonic dipole moment of a given harmonic order \textit{q} can be written as a Fourier sum over different quantum paths \textit{j} contributing to the XUV emission. Hence the quantum paths are a generalization of the semi-classical electron trajectories described in the three steps model. The time dependent dipole induced is given as follow:

\begin{align}
		d_q=\sum_j d_q^{(j)}\times e^{(-i\Phi_q^{(j)})}
\end{align}

where \textit{q} is the harmonic order, \textit{j} define the quantum path, $d_q^{(j)}$ corresponds to the dipole amplitude over a quantum-path \textit{j} and $\Phi_q^j$ is the respective dipole phase:

\begin{align}
	\Phi_q^j=I_p\tau_q^j+\int_{0}^{\tau_q^j} \frac{(\bf{p}+\bf{A})^2}{2} dt
\end{align}

where \textit{$I_{p}$} is the biding energy of the ground state (first ionisation potential level), $\tau_q^j$ is the excursion time also known as the duration of a quantum-path \textit{j} corresponding to the difference between ionisation times and recombination times, \textit{\bf{p}} is the canonical momenta and \textit{\bf{A}} the vector potential.
Because the ionisation time window differs from the recombination time window, high harmonics are not emitted at the same time \cite{mairesse_science_2003}. This lack of synchronisation, the well known attosecond chirp also present in the classical description, is actually an advantage for Attosecond spectroscopy as this way each harmonic order maps a given time in the laser field. This time scale depends on the laser field wavelength $\lambda_0$ and the laser peak intensity $I_0$, as the cut-off position scales with $\lambda_0^2$ and $I_0$, but also on the trajectories followed.
The mapping law in the plateau for each trajectories is given by the short and long trajectories excursion times (duration of each path). To estimate this law we performed quantum calculations of the harmonic dipole (under the Strong Field Approximation) and we derived an analytic expression of the EWP excursion time for the short and the long trajectories as a function of the harmonic order q  performing a fit (that for the short trajectory is consistent with the expression derived in M. Lein in \cite{lein_jphy_2007}):

\begin{widetext}
	\begin{equation}
\tau_{short}=\frac{\omega}{2\pi}[\frac{C_{1}}{\omega} (\frac{1}{\pi}\times arcos(1-\frac{2\omega(q-I_{p})}{\pi U_{p}C_{5}}))^{C_{3}}+\frac{C_{2}}{\omega} (\frac{1}{\pi}\times arcos(1-\frac{2\omega(q-I_{p})}{\pi U_{p}C_{5}}))^{C_{4}}]
\end{equation}
\begin{equation}
\tau_{long}=11.87-\frac{\omega}{2\pi}[\frac{C_{1}}{\omega} (\frac{1}{\pi}\times arcos(1-\frac{2\omega(q-I_{p})}{\pi U_{p}C_{5}}))^{C_{3}}+\frac{C_{2}}{\omega} (\frac{1}{\pi}\times arcos(1-\frac{2\omega(q-I_{p})}{\pi U_{p}C_{5}}))^{C_{4}}]
\end{equation}
\end{widetext}	

\begin{table}[here]
\centering
\topcaption{Constants for short and long excursion times calculation}
\begin{tabular}{@{} lcr @{} lcr @{} lcr @{} lcr @{} lcr @{} lcr }
\toprule
$        $ & $C_{1}$ & $C_{2}$ && $C_{3}$ & $C_{4}$ & & $C_{5}$ \\
\midrule
short& 2 & 2.1 & & 1.18 & 0.21 & & 1.06\\
long& 1.46 & 2.03 & & 1.53 & 0.16 & & 1.06\\
\bottomrule
\end{tabular}
\label{tab:booktabs}
\end{table}
where, $\omega$ is the fundamental laser frequency, $I_p$ the ionisation potential, $U_p$ the quiver energy, $q$ the harmonic order and $C_n$  the specify constants in Tab I. The results $\tau_{short}$ and $\tau_{long}$ are given in optical cycle units.

Under our experimental conditions ($800$ nm and $I_0=0.5-2 \times 10^{14} W.cm^{-2}$), we present the mapping law in figure 2. As one can see depending on the laser peak intensity $I_0$ (colorbar) the cutoff extends toward higher harmonic order as expected. For the maximum extension (we employed a maximum peak intensity of $2\times 10^{14} W/cm^{2}$) the short trajectory maps a time scale from $\sim 200$ asec to the position of the excursion time of the cutoff $1.7$ fsec whereas the long trajectory maps a time scale from the cutoff excursion time ($1.7$ fsec) to $\sim 2.7$ fsec. 
Thus, if any intra-cation dynamics occurs within this time window, the total dipole moment is modified dictating that the HHG radiation encodes key signatures of this motion just after ionisation \cite{ lein_jphy_2007, marangos_pchem_2008}. We will indeed examine the case of fast nuclear motion, channel superposition signature and channels dynamical populations transfer due to laser coupling that happen during this time scale.
\begin{figure}[here]
\centering
\includegraphics [width=3.5 in,height=2.5 in]{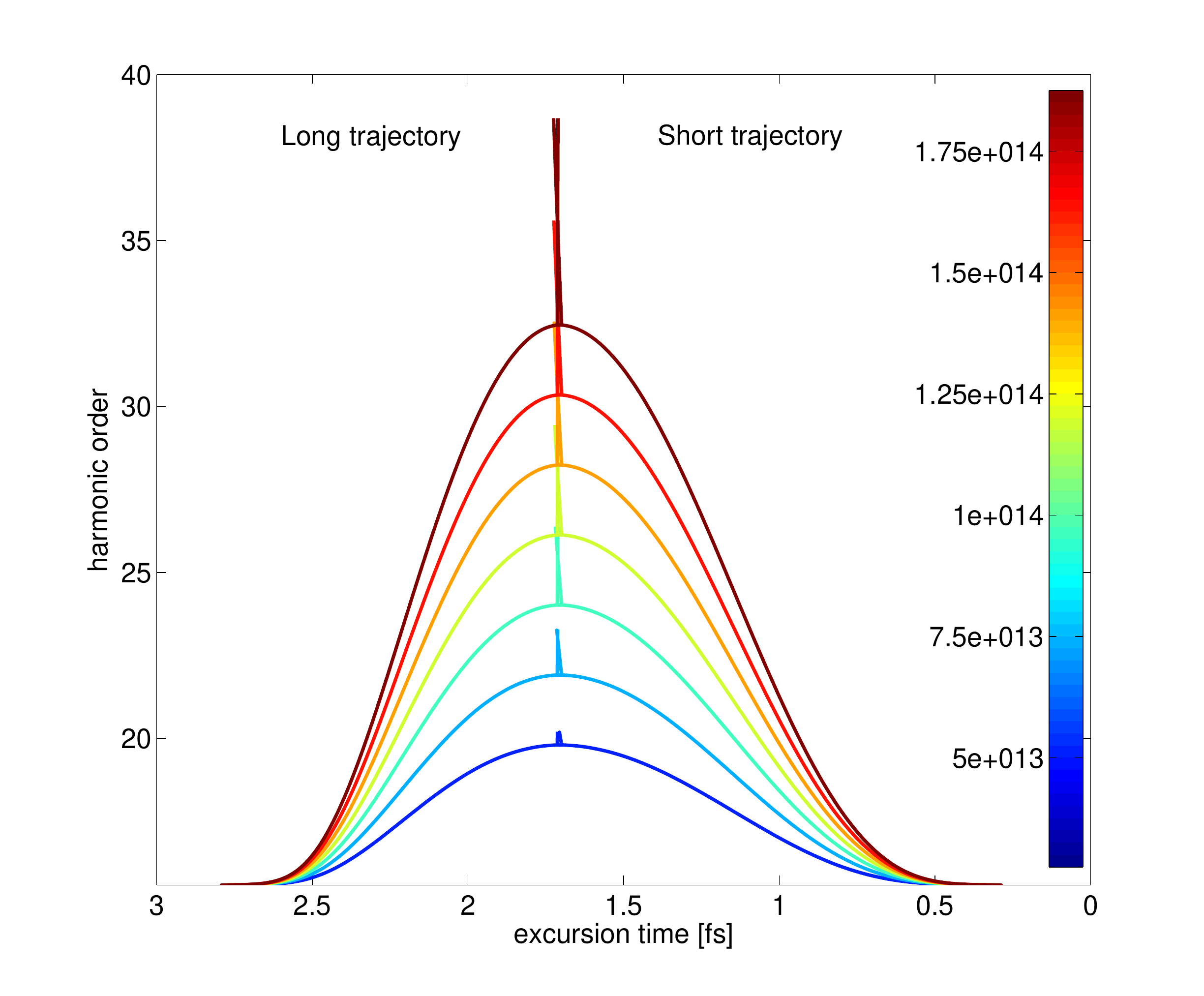}
\caption{Excursion time calculated for each trajectories within laser peak intensity of $I_0=0.5-2 \times 10^{14} W.cm^{-2}$ as employed in the experiment and shown in the colour bar. The mapping law for both trajectories allows us to follow dynamical process from $200$ asec to $1.7$ fsec for the short trajectory and from $1.7$ fsec to $2.7$ fsec for the long trajectory}
\label{spatial_short_long}
\end{figure}

To access the encoded information, experimental methods based on HHG have been employed. The PACER technique (Probing Attosecond motion by Chirp Encoded Recollision), that accesses the cation dynamics via the time-frequency mapping allowed by the attosecond chirp, was first implemented for retrieving nuclear dynamics by comparing the HHG spectrum between deuterated (reference) and protenated molecules of the same species \cite{baker_science_2006,marangos_pchem_2008}. The tomography technique \cite{itatani_nature_2008, hassler_nphys_2010, diveki_nphys_2012} permits to access information on which ground states (molecular orbitals) is involved in the HHG process and for this a normalisation to an atomic partner response (reference) is required. Recent works \cite{mairesse_prl_2010} based on two HHG source interferometry, one source in unaligned molecules (reference) and the other one in aligned molecules, shows that the HHG signal contains crucial information on the electronic orbitals involved in the process and their fast dynamics. The ideal tool would be to access these dynamics without recourse to an auxiliary reference to extend the possibility of such pump-probe scheme to reveal in a more versatile way the dynamical information.

One route to this self-referenced Attosecond Spectroscopy is to take advantage of the two quantum-paths (short and long) and especially of their interferences (self-referenced technique referred to as QPI for quantum path interferences). Their corresponding ionisation and recombination times being specific to a quantum-path,  the high harmonics emission time are separated by 100's of asec and map the full time scale needed to study dynamical processes. Note that if in the presented results we use 800 nm laser field  and make a proof of principle of QPI for Attospectroscopy, one way to extend even more the time scale of study (see eq.3 and 4) will be to take advantages of NIR sources that provide wavelengths (1200-2000nm) higher than 800 nm. Hence the molecular QPI technique can be extended to more complex molecules where dynamics due to ionisation and subsequent structural changes will occur beyond 2-3 fsec. 

Preliminary studies on interferences of electron trajectories involved in the HHG process (know as atomic Quantum-Path Interferences QPI) has demonstrated the possibility to access experimentally dipole information in atoms \cite {zair_prl_2008,auguste_pra_2009}. 
As shown on figure 3, one can see how the harmonic yield is modulated corresponding to subsequent constructive and destructive interferences of the quantum paths with respect to the laser peak intensity. The experimental data shown in red are consistent with the SFA calculations (blue) whereas the TDSE calculation are not showing clear modulations with a high enough contrast due to the non-selection of the two first quantum-paths (short and long) as allowed within the SFA model. The periodicity of the modulation close to $0.3\times 10^{14} W.cm^{-1}$ is a good indication that the interferences pattern is due to short and long trajectories interferences and not higher order trajectories (multiple return). 

 \begin{figure}[here]
\centering
	\includegraphics [width=3 in,height=2 in]{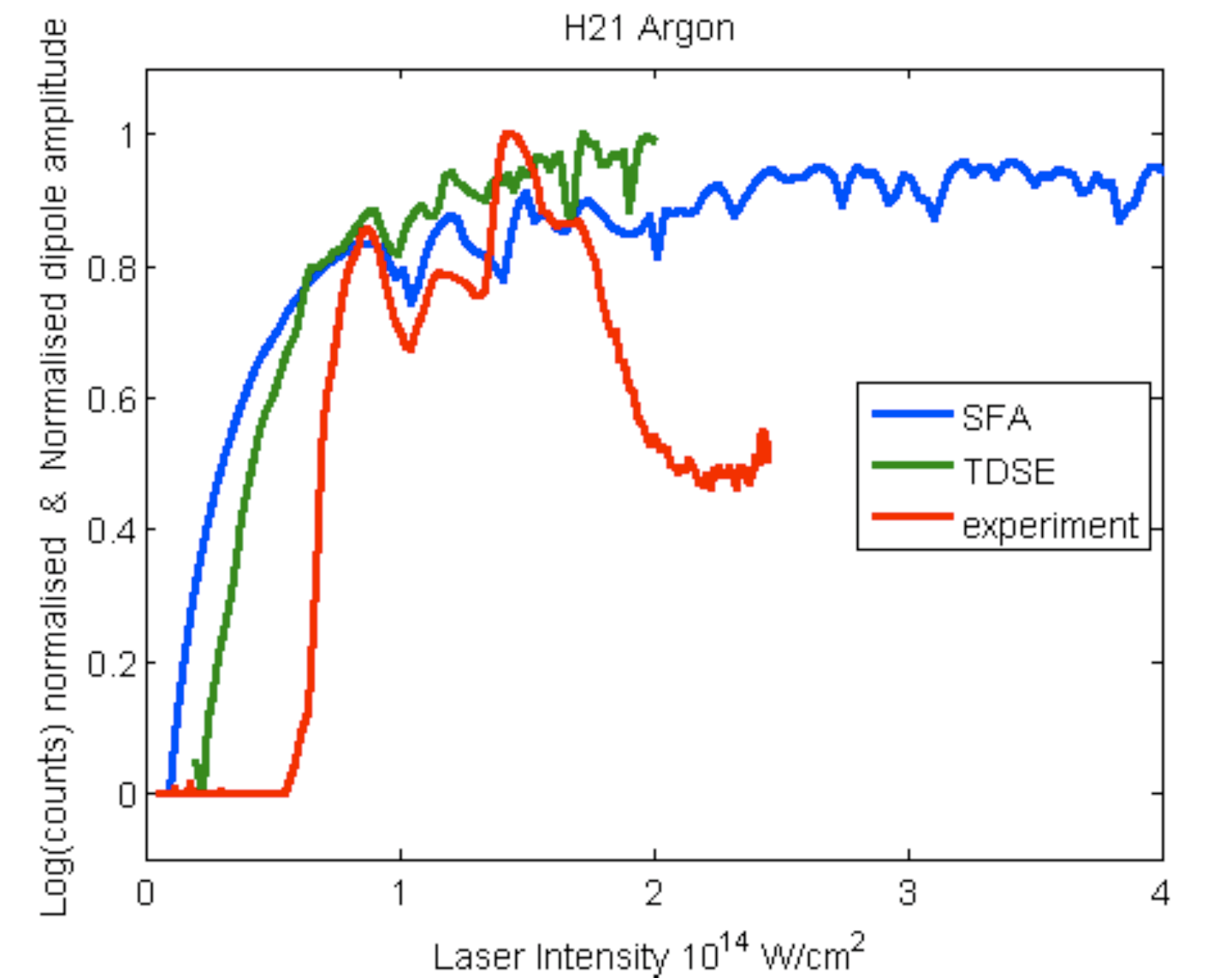}
\caption{Quantum path interference calculated with TDSE (green), SFA (blue) and measured experimentally in Argon jet target (red) for harmonic $21^{st}$.}
	\label{fig:fig3}
\end{figure}

These studies demonstrated that experimental conditions can be found where the short and long trajectories may interfere, creating  an 'interferometer' with a resolution of 10's asec. The interferometer arms are experimentally controllable in phase and amplitude. The phase of each arm is proportional to the ponderomotive energy $U_{p}$ (i. e. depends upon the laser peak intensity or the laser wavelength $U_{p}\sim I\times\lambda^{2}$) and the excursion time of the electronic wave packet in the continuum specific to an harmonic order and a trajectory. Therefore one way to control the relative phase of the interferometer arms is by controlling the laser intensity. The relative amplitude of the short and long trajectories that contribute to the macroscopic HHG signal can be controlled by: adapting phase matching conditions, controlling the focusing condition of the laser in the gas jet \cite{salieres_science_2001} or  using a second harmonic of the laser field \cite{brugnera_prl_2011}. 

Using the phase matching control procedure, both trajectories can be phase-matched together with different divergences which provides a direct spatial mapping of the trajectory contributions in the far field.
The divergence of the short trajectory contribution is smaller than the divergence of the long trajectory contribution. So the short trajectory appears mainly on-axis whilst the long trajectory appears off-axis. In the area where both contribution overlap their interference leads as spatial rings forming in the far-field intensity distribution. A far-field spatial selection was used to allow the balance of the two contributions to maximise the QPI contrast \cite{zair_prl_2008, auguste_pra_2009}.
This interferometric measurement offers a high sensitivity to any change in the cation which make it suitable for the investigation of ultra-fast intra-molecular dynamics. Importantly QPI is a self-referencing method since it uses the observation of the spatial and the spectral distribution of the harmonic emission in a single experiment to extract the information.

In this article we show for the first time, an extension of the QPI to the case of molecules in order to retrieve electronic and nuclear dynamical cation information.

\section{Experimental investigation of nuclear motion }
All experimental results presented here were performed using the same configuration: we used a Ti:sapphire CPA laser system (Red Dragon, KML Inc.) delivering 30 fsec, 800 nm pulses at a repetition rate of 1 kHz with a maximum pulse energy of 8.5 mJ of which we used  3 mJ in the full beam before aperturing for our experiment. The laser beam entered a vacuum chamber and was focused by a 40 cm radius of curvature (ROC) spherical mirror into a pulsed jet of $500 \mu m$ diameter aperture with 1 bar backing pressure. A flat field spectrometer, composed of a grazing incidence grating ($87^{\circ}$, 1200 gr/mm, ROC=200 mm) was used  to spectrally disperse the HHG radiation in one direction whilst preserving the divergence in the perpendicular direction. 
A multi-channel plate (40 mm, Photonis Inc.) was placed in the grating focal plan followed by a backside-illuminated CCD, to allow us to measure the spatially and spectrally resolved harmonic emission. The acceptance angle of our detector, 20 mrad, was sufficient to resolve spatially the short and the long trajectories. The gas jet was positioned $\sim2\,mm$ before the laser focus to phase-match both trajectories where the laser beam cross-section diameter was $75 \mu m$.
An aperture at the entrance of the HHG chamber was used to optimise the signal from the distinct quantum paths and the QPI, in this case the laser beam was apertured down to 4 mm so that the maximum energy throughput is $\sim 0.6 mJ$ for all the experiment. A variable attenuator (wave plate/polariser) was placed in the generating  beam to control the laser peak intensity at the position of the jet and thus control the relative phase of the two trajectories. The acquisition of each harmonic spectrum was performed over 10 000 laser shots.

Randomly aligned molecules were investigated to demonstrate a first experimental evidence of the different dependence of the two quantum path contributions upon the fast molecular nuclear motion. Whilst the experiments are readily extended to aligned molecules, the purpose of the present study was to show a first proof of principle in the capability of the long and short trajectories to encode dynamical effects in the HHG signal recorded. The nuclear dynamics are expected to change the HHG emission amplitude through the evolution in time of the nuclear autocorrelation function \cite{lein_prl_2005, serguei_prl_2009}. We study respectively the case of $CO_{2}$ and $N_{2}$ that are expected to have a slow evolution of the autocorrelation function \cite {serguei_prl_2009}, the case of $O_{2}$ (expected to have a intermediate evolution of the autocorrelation function) and the case $D_{2}$ and $H_{2}$ (expected to have a fast evolution of the autocorrelation function but different from each other).
The accessible time window of our experiment is given by the difference of excursion time of the electronic wave packet in the continuum between the short and long trajectories. 
\begin{figure}
\centering
\includegraphics [width=3 in,height=4.25 in]{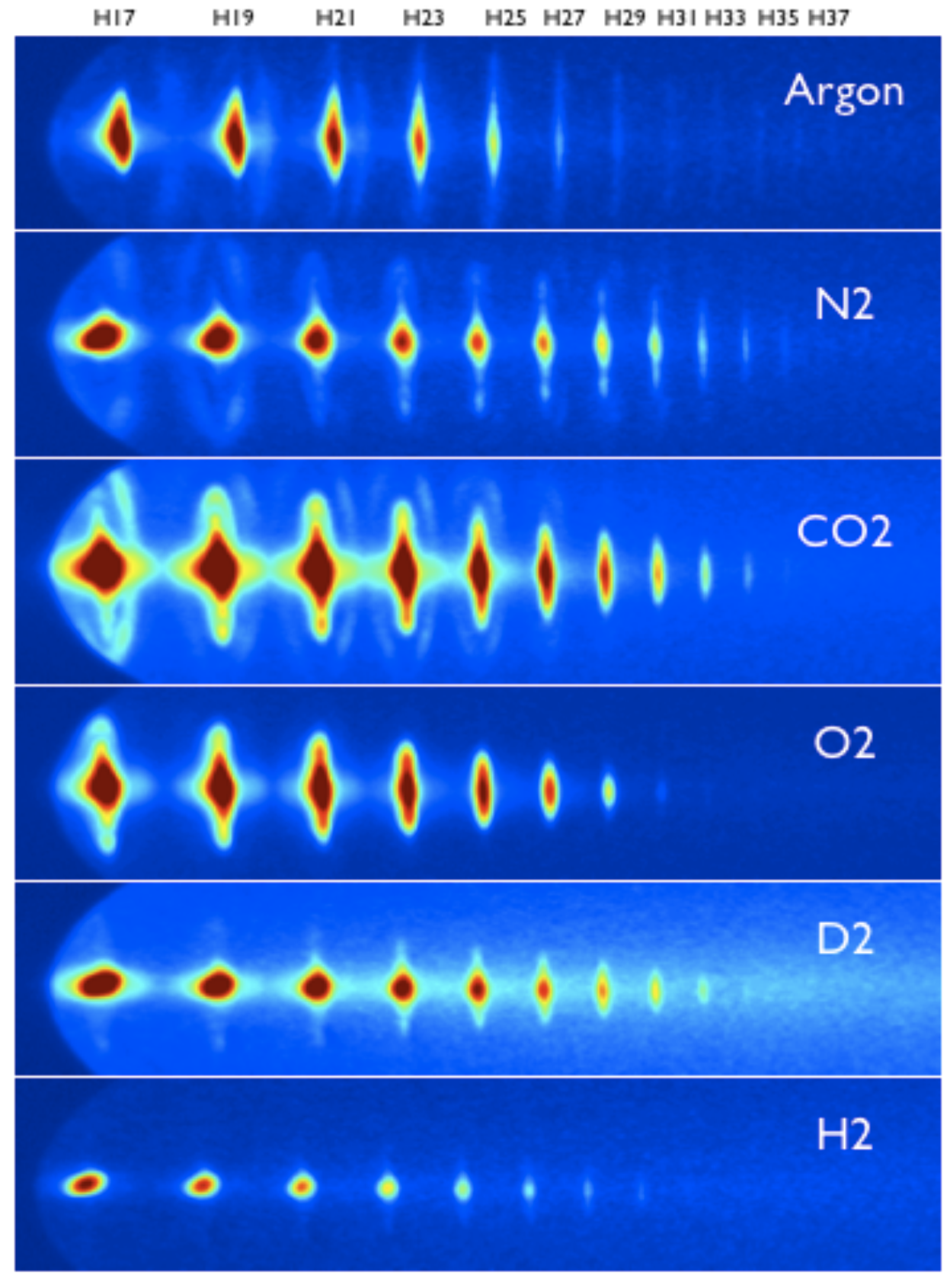}
\caption{Far Field spatial distribution of short (on axis) and long (off-axis) trajectories generated in Argon (upper) and in randomly aligned molecules ($N_{2}$,$CO_{2}$,$O_{2}$,$D_{2}$,$H_{2}$). We use for the HHG a 30 fs IR laser pulse at $2\times10^{14} W/ cm^{2}$ at the jet position located at  $\sim$ 2 mm before the laser focus and the backing pressure is kept at $1\,bar$. The horizontal axis corresponds to the spectral domain, the vertical axis corresponds to the harmonic beam divergence and the colour map is the harmonics signal in logarithmic scale. The elliptical 'shadow' that cut slightly the $17^{th}$ order on the left hand side is the edge of the MCP detector.}
\label{spatial_short_long}
\end{figure}

In Fig.4 HHG is shown in the far field for Argon, $CO_{2}$, $N_{2}$, $O_{2}$, $H_{2}$ and $D_{2}$  for an intensity at the position of the jet of $2 \times 10^{14} W/cm^{2}$. For the atom and most of the molecules, each harmonic in the plateau presents a similar shape: a central peak surrounded by an annular distribution. These HHG spatial distributions correspond to the signature of short trajectory mainly phase-matched on-axis (central spot) and long trajectory mainly phase-matched off-axis (ring). The inner spatial ring observed around the on-axis region, well resolved in the case of $N_{2}$ and $CO_{2}$, corresponds to the region of spatial interferences of the two trajectories. All measurements were made maintaining the same geometry, the same backing pressure and the same jet position, allowing comparison between all the samples. Figure 4 shows a clear difference of the recorded signal from one molecular target to the other.
The HHG signal of $N_{2}$ and $CO_{2}$ implies that if the nuclear motion is slow for these molecules, both trajectories 'see' the same cation at ionisation with minimal changes  by the instant of recombination.
In the case of $D_{2}$ and $H_{2}$ the results obtained are highly interesting. Actually being deuterated and protenated partners, these two molecules illustrate the most the signature of the nuclear motion differences encoded into the spatially and spectrally resolved short and long trajectory contributions. As one can see the HHG signal is highly affected, indeed the long trajectory contribution is clearly diminished, being almost totally suppressed in the $H_{2}$ case while still present in $D_{2}$. 
This is consistent with the idea that the intensity of harmonic radiation is depending upon the nuclear motion and is modulated by the square modulus of the nuclear autocorrelation function given by $|<\psi_{n}(t)|\psi_{c}(t)>|^{2}$, where $\psi_{n}$ (respectively $\psi_{c}$) is the wave packet evolution of the neutral (respectively of the cation). Due to the presence of the short and long trajectories, for a given harmonic order, it is then possible to follow this nuclear dynamics over broad time scale, here 200 asec to 2.7 fsec. The autocorrelation function can be calculated for various molecules as nicely shown in \cite{serguei_prl_2009}.  For fast nuclear dynamic as for $D_{2}$, $H_{2}$ the autocorrelation function decrease rapidly over time compare to $O_{2}$  and  $CO_{2}$ and $N_{2}$ having a autocorrelation function evolving slowly (autocorrelation function almost constant over this time scale). Hence our observation in $D_2$ and $H_2$  is a clear experimental demonstration that the faster the molecule expand the more affected the HHG signal from the long trajectory contribution. The case of $O_{2}$ appears to be intermediate to ${N_{2}; CO_{2}}$ and ${D_{2}; H_{2}}$. Even though these molecules do not have the same structure and so won't show the same harmonic yield, it illustrates plainly the fact that the relative strength of short and long trajectories is sensitive to nuclear motion of the cation with attosecond-femtosecond time scale, especially in $H_{2}$ and $D_{2}$ where the comparison is straight.

\begin{figure}
\centering
\includegraphics [width=3.5 in,height=2.5 in]{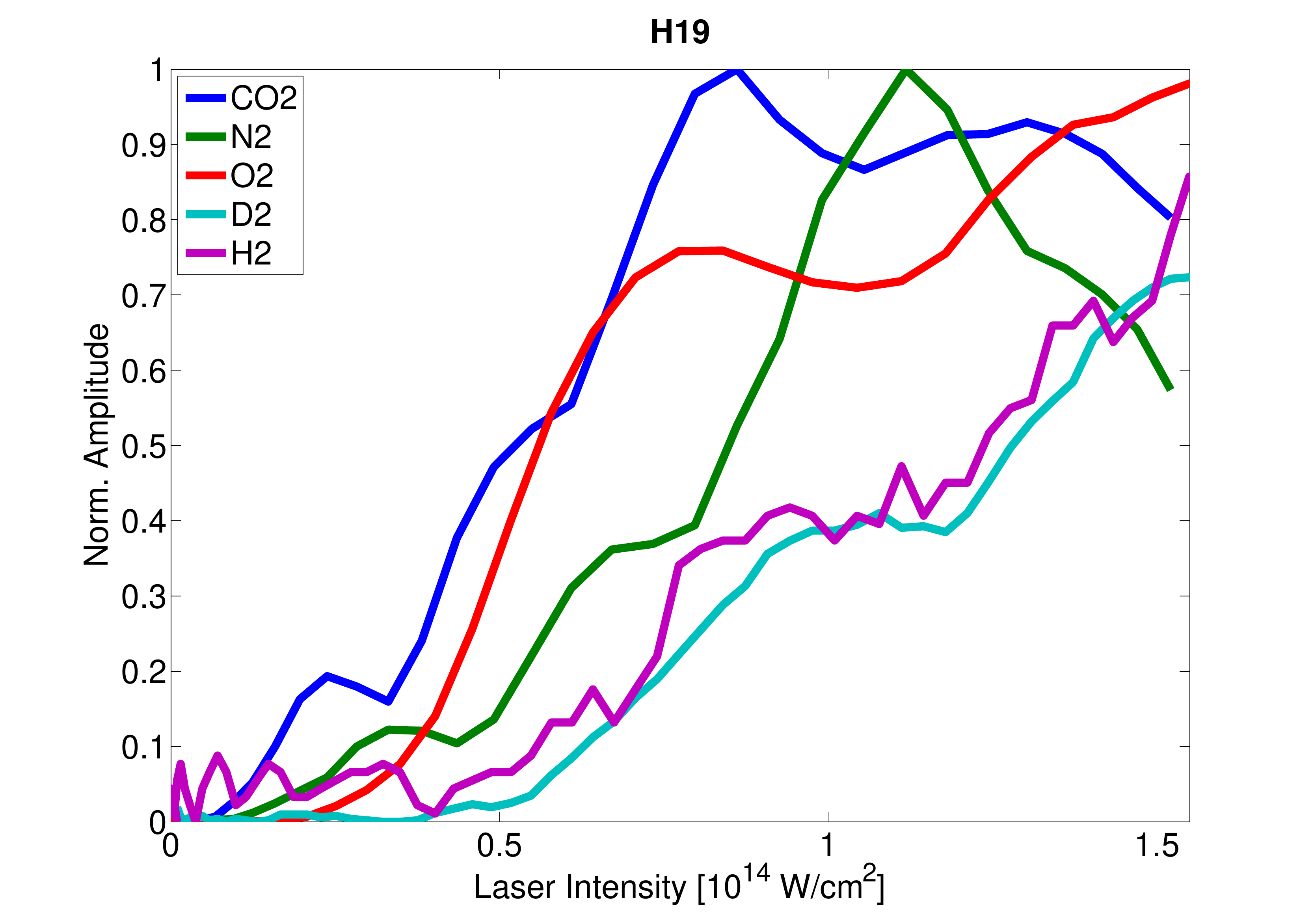}
\caption{Far Field spectrally resolved QPI of 19th  for the following unaligned molecules: $CO_{2}$, $N_{2}$, $O_{2}$, . In the case of$D_{2}$ and $H_{2}$ we can observe the disappearance of the modulation which shows that the long trajectory is not contributing the QPI pattern.}
\label{fig:example}
\end{figure}

To investigate further the role of cation nuclear dynamics we performed an intensity scan of the HHG spatial distribution.  The intensity was varied from 0.5 to $1.5\times10^{14}\,W/ cm^{2}$ using an attenuator. While changing the intensity, the annular distribution evolved from the on-axis peak to the annular off-axis distribution.  To obtain spectrally resolved Molecular QPI, we integrate the signal over a spatial region located from 10 to 12\,mrad  from the central axis (where interference between the short and long contributions is the strongest), for all the molecules and record the harmonic signal at the harmonic frequency as a function of the laser peak intensity. The signal obtained presented in figure 5 for all the molecular samples, shows a fringe contrast that differs between the molecules due to the fact that the strength of the long trajectory is dependent on the molecules. However the periodicity seems to be conserved at $\sim 0.3-0.4 \times 10^{14} W/cm^{2}$, consistent with previous observations in atoms \cite{zair_prl_2008, auguste_pra_2009}. This confirms that SFA calculation, valid for atomic case, can be also use to investigate how the QPI are evolving while generating harmonic in unaligned or aligned molecules.

Thus the difference in the long trajectory amplitude observed between each molecular sample can only be attributed to nuclear motion in the cation between ionisation and recombination. The Molecular QPI modulation periodicity appears to be unchanged in all molecules studied here indicating the robustness of the continuum electron behaviour and setting the conditions for the detection of different cation evolution.
The observation of short and long trajectories in a broad range of molecules (Fig.4) and the confirmation of QPI signatures (Fig.4 and 5) suggests the use of molecular QPI  to quantify cation dynamics. We now consider theoretically how electronic dynamics should be accessible from such Molecular QPI measurement.

\section{Theoretical investigation of molecular QPI and ionisation channel dynamics}

We know from previous studies \cite{brian_science_2008, murnane_chemphys_2009} that in $N_{2}$ the parallel alignment favours the contribution to HHG from the HOMO and HOMO-2 orbitals (large dipole due to orbital symmetry) whereas the contribution of HOMO-1 is cancelled (no dipole: nether ionisation nor recombination is possible due to orbital symmetry). In perpendicular alignment the HOMO-1 is favoured (large dipole due to orbital symmetry) and HOMO and HOMO-2 are diminished (small dipole: ionisation and recombination are smaller than in parallel alignment) but not entirely cancelled. For a given laser intensity, the HOMO dominates up to the position of its cutoff and the contribution of HOMO-1 is then revealed beyond the HOMO cutoff where it dominates the signal over HOMO-2 contribution. However in the case of $N_{2}$ the difference of ionisation potential for HOMO and HOMO-1 is only one photon energy at 800 nm, therefore tracking a signature of this orbital involved in the extended cut-off while rotating the molecule is challenging. Additionally this separation of the ionisation channels by just one photon energy (800 nm) may allowed a transfer of electronic population by laser coupling changing all the orbitals contribution weight upon few 100's of asec. 

It is needed to find methods not only based on cut-off extension of the harmonic spectra to explore this intra-cation dynamics. Proving that the Molecular QPI are sensitive to nuclear dynamics of the cation, we will in this section exploring the capability of Molecular QPI for probing multiple orbitals and molecular cation electronic dynamics.
We first investigated theoretically the Molecular QPI signal dependence upon different scenarios for $N_{2}$ like and model $H_{2}$ like molecules.

\subsection{Atomic model for QPI probing ionisation from multiple channels}
Our first approach is very simple and consists in understanding how the QPI measurement may be sensitive to the ionisation channels. For this we have calculated the QPI pattern for a Hydrogen-like atom but with the ionisation level of $N_{2}$ orbitals: HOMO ($I_{p}=15.58 eV$) named channel 1, HOMO-1 ($I_{p}=16.93 eV$) named channel 2 and HOMO-2 ($I_{p}= 18.78 eV$) named channel 3. For this we use the Strong Field Approximation and calculate the dipole over a multi-cycle laser pulse close to experimental conditions ($2\times10^{14}\,W/ cm^{2}$; Gaussian envelop; 30 fs FWHM). The strongest contributions being the short and the long trajectories at the high harmonic frequency. The results is shown in figure 6 where clear QPI modulations are resolved for the $19^{th}$ harmonic H19 and for which the periodicity is found to be $\approx0.3\times10^{14}\,W/ cm^{2}$. Similar modulations are found for other harmonics order. As one can see  the QPI pattern presents a clear shift for the different channels used, this shift increasing toward  lower intensity with the $I_{p}$ level. If we consider that the QPI pattern, showing subsequent maxima that correspond to constructive interferences between short and long trajectories, follows a modulated function of $2\pi$ periodicity hence the channel 2 is shifted by $\frac{\pi}{2}[2\pi]$ with respect to channel 1 and $\pi[2\pi]$ for channel 3 with respect to channel 1. This is due to the term $I_{p}^{orbital}\tau_{(j)}$ in the phase acquired by the electronic wave packet upon a specific trajectory $(j)$ , tunnelling out and recombining to a given orbital (single channel) \cite{lewenstein_pra_1995}. Increasing the $I_{p}$ also induces a decrease of the recombination and ionisation probability therefore as observed the QPI yield is lower for channel 3. 
In this very simple approach no orbital symmetry is taken into account nor laser coupling,  rotating the molecule would correspond to an averaging over the QPI patterns of channel 1 (HOMO) and channel 2 (HOMO-1). Therefore one of the signatures of multiple orbitals involved in the HHG process onto the QPI signal would be mainly a shift in the intensity position of the maxima and minima.
This atomic-like' approach shows even in a very simple manner that if many orbitals are involved their signature should be encoded in the Molecular QPI.

\begin{figure}
\centering
\includegraphics [width=3.5 in,height=2 in]{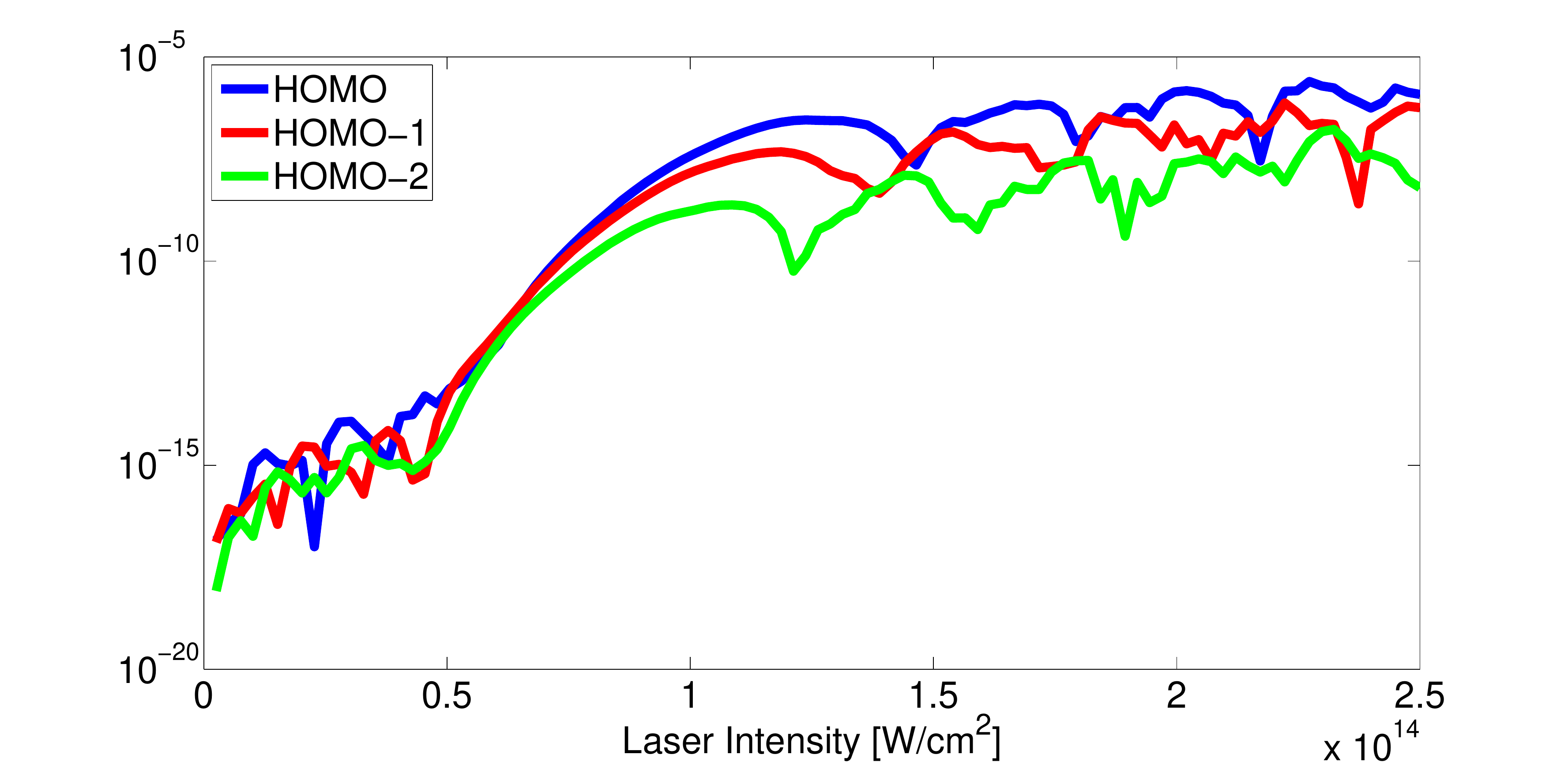}
\caption{Atom-like QPI pattern of 19 th harmonic calculated for a Hydrogen-like molecule with 1s symmetry and $N_{2}$ ionisation level of its HOMO, HOMO-1 and HOMO-2}
\end{figure}

\subsection{Molecular model for QPI probing dynamical transfer of population between ionisation channels}

During the HHG process, QPI should have the potential to separate different scenarios: In aligned $N_{2}$ molecules, single channel  starting from HOMO or HOMO-2 orbitals, and their superposition may contribute to the HHG. In perpendicularly aligned $N_{2}$ molecules, the contribution of HOMO-1 arise and the contribution of HOMO-2 can be considered negligible (less that 5\% of the population is in this state \cite{mairesse_prl_2010}). In this case an additional scenario may be considered where a transfer of population between HOMO and HOMO-1 induced by the laser field is allowed. In a one colour laser field at 800 nm, this transfer of population is expected to be more favourable in $N_{2}$ than in $CO_{2}$ due to the energy gap between these channels (one photon energy for $N_{2}$; three photon energy for $CO_{2}$). Therefore in a first  quasi-analytical approach we present a study of the QPI in $N_{2}$ molecule with respect to its alignment to the generating laser field.  To go even deeper in the understanding of such signature we present in a second study an ab-initio calculation of a model $H_{2}$ molecule subject to an additional resonant field that induces this transfer of population and allowed to study more accurately its signature in QPI pattern of each separated channels.

\subsubsection {Semi-analytical model for High-Harmonic Generation in molecular Nitrogen}
 As beautifully demonstrated  by Smirnova et al \cite {SmirnovaProcNatlAcadSci_2009, olga_nature_2009},  nitrogen cation (at a fixed internuclear distance taken at equilibrium) has two close ionic states,  ionisation from HOMO (called X) and ionisation from HOMO-1(called A) that can be pictured with Hartree-Fock method as:
 \begin{equation}
\mid1> \equiv X   ^{2}\Sigma^{+}_{g}:
(1\sigma_{g})^{2}(1\sigma_{u})^{2}(2\sigma_{g})^{2}(2\sigma_{u})^{2}(1\pi_{u})^{4}(3\sigma_{g})
\end{equation}

\begin{equation}
\mid2>\equiv A   ^{2}\Pi^{+}_{u}:
(1\sigma_{g})^{2}(1\sigma_{u})^{2}(2\sigma_{g})^{2}(2\sigma_{u})^{2}(1\pi_{u})^{3}(3\sigma_{g})
\end{equation}

In the laser field, this two level system ($\mid1>$ and $\mid2>$ with $E_{1}$ and $E_{2}$ energies) is split in two
quasi-static states (i.e. polarised states) $\mid\tilde1>$ and $\mid\tilde2>$ with $\epsilon_{1}$ and $\epsilon_{2}$ energies. In the field free basis and at the ionisation time $t_{i}$, these two states can be written as:
\begin{equation}
\begin{pmatrix} \mid\tilde1>\\\mid\tilde2>
\end{pmatrix}
= \frac{1}{\sqrt{\Delta^{2}(t_{i})+V^{2}(t_{i})}}
\begin{pmatrix} \Delta(t_{i}) & -V(t_{i})\\V(t_{i}) & \Delta(t_{i})
\end{pmatrix}
\begin{pmatrix} \mid1>\\\mid2>
\end{pmatrix}
\end{equation}
Where $\Delta(t_{i})=E_{2}-\epsilon_{1}(t_{i})=-(E_{1}-\epsilon_{2}(t_{i}))>0$.
At $t=t_{i}$ the molecule is in a superposition of these two quasi-static states: $\Psi_{i}=\tilde a_{1}(t_{i})\mid\tilde1>+\tilde a_{2}(t_{i})\mid\tilde2>$. By solving the Time Dependent Schr\"odinger Equation (TDSE) we can obtain the evolution of the  two quasi-static level system in the field free basis.
Taking the molecular state after ionisation, we can calculate the dipole along each trajectories ($traj=$ short or long) and for each channel ($\tilde c=\tilde 1$ or $\tilde 2$) using SFA calculations.
Using the saddle point approximation as in \cite{ivanov_pra_1996}, the Fourier transform of the dipole can be written as :
\begin{equation}
d(\Omega) = \sum_{t_i,t_r}{A_{ion}(ti) A_{prop}(ti, tr) A_{rec}(ti, tr)}
\end{equation} 
where the ionisation time $t_i$ and the recombination time $t_r$ are dependent on the frequency $\Omega$ corresponding to harmonic energy. Hence:
\begin{widetext}
\begin{equation}
d_{traj}^{\mid\tilde c>} (\Omega)=A_{ion}^{\mid\tilde c>}(t_{i,traj})A_{prop}(t_{i,traj},t_{r,traj})
\times [a_{1}^{\mid\tilde c>}A_{rec}^{\mid1>}(t_{i,traj},t_{r,traj})+a_{2}^{\mid\tilde c>}A_{rec}^{\mid2>}(t_{i,traj},t_{r,traj})]
\end{equation}
\end{widetext}
where,
$A_{ion}(t_{i})$ is the ionisation amplitude at time of ionisation $t_{i}$, we use the ADK ionisation rate 
$\exp{-\frac{1}{3}\frac{(2I_{p})^{3/2}}{E_{L}(t_{i})}}$.
The propagation amplitude between the ionisation is given by: 
\begin{equation}
A_{prop}(t_{i}, t_{r})=\frac{1}{(t_{r}-t_{i})^{3/2}} e^{-i \int_{t_{i}}^{t_{r}} dt' \xi(t')-iI_{p}(t_{i}-t_{r})}\\
\times e^{i\Omega t_{r}}
\end{equation}
And $A_{rec}(t_{i}, t_{r})=<g\mid p\mid \Psi_{ion}\Psi_{k}(t_{i},t_{r}) >$ the recombination amplitude at time of recollision $t_{r}$ between the ground state $\mid g>$ and the exited system ( $\Psi_{k}(t_{i},t_{r})$ being the continuum state and $\Psi_{ion}$ the ionic state); $e^{-i \int_{t_{i}}^{t_{r}} dt' \xi(t')}$ is corresponding to the Volkov propagator with $\xi(t')= \frac{1}{2}p(t')^2$.


To take into account the difference of symmetry between HOMO (even function along the electron coordinate; x-axis) and HOMO-1 (odd function along the electron coordinate) and the symmetry of the transition (without making any assumption on the dipole) we use the plane wave approximation of the continuum state $\exp(ikx)$ and if we define $\Psi^{H}$ for HOMO and $\Psi^{H-1}$ for HOMO-1 as the spatial wave-functions of the ionic states then:
\begin{widetext}
\begin{equation}
A_{rec}^{\mid1>}=\int dx \Psi^{H}(x,y,z) (-i \frac{\delta}{\delta x}) \exp(ikx)
=k.FT_x (\Psi^{H})
=k \int dx \Psi{H}(x,y,z) cos(kx)
=-1 \times f_1(k(t_i))
\end{equation}

\begin{equation}
A_{rec}^{\mid2>}=\int dx \Psi^{H-1}(x,y,z) (-i \frac{\delta}{\delta x}) \exp(ikx)
=k.FT_x (\Psi^{H-1})
=i k \int dx \Psi{H-1}(x,y,z) sin(kx)
=i \times f_2(k(t_i))
\end{equation}
\end{widetext}

with $f_1(k)$ being an even function and $f_2(k)$ being an odd function. Both functions  are set positive for $k<0$ (the sign of k changing together with the direction of laser field). Therefore to represent the even and odd characteristics of $f_1$ and $f_2$, we have taken: $f_1(k(t_i))=1$ if  the laser field $E(t)>0$, else $f_1(k(t_i))=-1$; and $f_2(k(t_i))=1$.

For the ionisation amplitude,  we need to take into account the HOMO and HOMO-1 orbital coordinates from which we can determine what is the sign of the orbital lobe in the direction of the ionisation to add a $\pi$ shift to the ionisation amplitude. The Amplitude of  $\mid \tilde 1>$ and $\mid \tilde2>$ are both negative if we ionised from positive coordinates. If we ionise form negative coordinate, the amplitude for $\mid \tilde 1>$ gets  negative and the amplitude for $\mid \tilde 2>$ gets positive.
To represent this coordinates effect we choose: $A^{\mid \tilde 1>}_{ion}(t_i) = -1 \times A^{\mid \tilde 1>}_{ADK}(t_i)$ 
;
$A^{\mid \tilde 2>}_{ion}(t_i) = 1 \times A^{\mid \tilde 2>}_{ADK}(t_i)$, if $E(t)>0$\\
$A^{\mid \tilde 2>}_{ion}(t_i) = -1 \times A^{\mid \tilde 2>}_{ADK}(t_i)$, 
if $E(t)<0$\\


The transition between the two quasi-static ionic states is only permitted in perpendicular configuration of the laser field versus the molecular internuclear axis (angle $\alpha$) due to molecular orbitals symmetry in $N_{2}$. In other words, the transition dipole being maximum ($\sim 0.25 a.u.$ \cite{mairesse_prl_2010}) for molecules perpendicular to the laser field and null for molecules parallel to the field, the parallel component of the coupling is null and the perpendicular component is given by:  $V (t_{i},\alpha)= \bf{-d.E_l}$ $=0.25 E_{L}(t_{i})\times sin(\alpha)$).
From this semi-analytic approach we calculate the QPI pattern as a function of the alignment angle. Results are shown in figure 6 for the harmonic $21^{st}$and calculating over one single laser period to reduce the calculation to the short and the long trajectories only. 

\begin{figure}
\centering
\includegraphics [width=3 in,height=2.5 in]{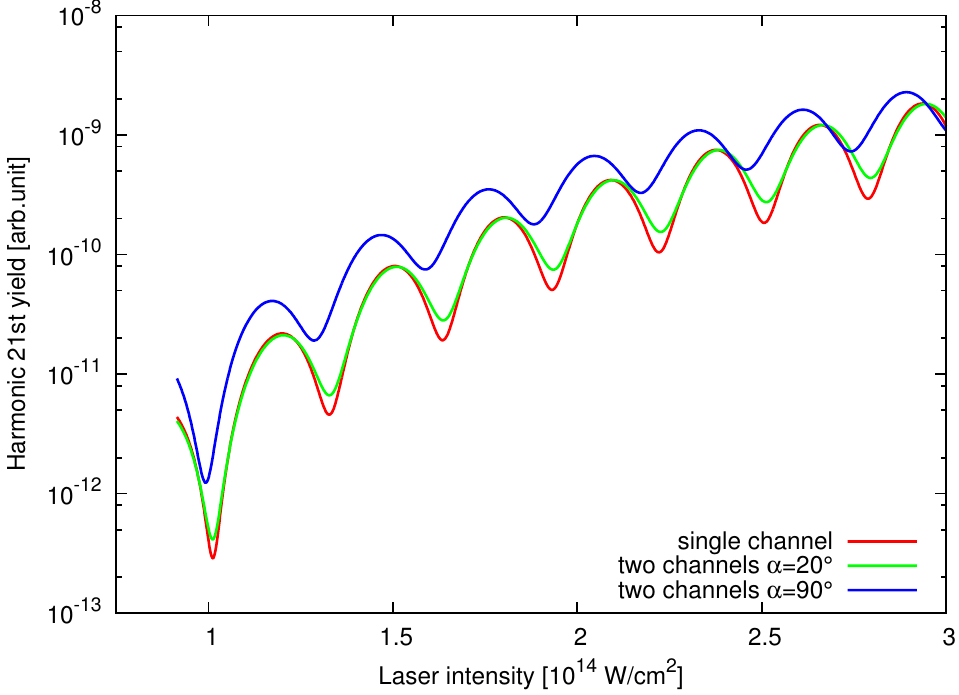}
\caption{Molecular  QPI pattern for 19th harmonic calculated for $N_{2}$ as a function of the alignment angle $\alpha$. The contribution of the two channels HOMO  and HOMO-1 only differ from the single HOMO channel contribution while the molecule is perpendicular to the laser}
\label{figure 6}
\end{figure}

In the case of a single channel shown in the plain blue curve, the QPI modulation are nicely reproduced with a periodicity of $\approx 0.3\times 10^{14} W/cm^{2}$ as expected. At an alignment angle of $\alpha=20\circ$ corresponding to the acceptance angle achieved experimentally, the two channel QPI pattern, shown in dashed red line, does not differ from the single channel one confirming the dominant contribution from HOMO (channel X) in the resulting HHG yield. However in the case of $\alpha= 90\circ$ which corresponds to a perpendicular alignment of the molecules with respect to the laser field, the QPI pattern is clearly shifted by $\pi/2$ towards the lower intensity as for the atomic model. However, although the periodicity is conserved, the contrast of the QPI is also affected. Therefore we think that the shift is a clear signature of the multiple channel contribution and that the contrast is a key element to identify the laser coupling population transfer between HOMO and HOMO-1.

\subsubsection {ab-initio TDSE calculation for two electron system }

In order to understand how the laser coupling could be disentangled from the multiple channels contribution we have performed a more sophisticate calculation based on full quantum mechanical calculation on a 2D two-electrons model system based on Multi-Configuration Time Dependent Hartree method (MCTDH) \cite{MCTDH}. The method implementation details for the strong field phenomena can be found in \cite{suren_pra}. This system corresponds to a model molecule with fixed internuclear distance, R = 3.54 a.u., having a parallel to the molecular axis coupling between HOMO and HOMO-1 with an energy gap of 3 photon energy form the laser fundamental frequency.
To emphasis the implication of the laser coupling states compare to the contribution of multiple channel, we considered the case of molecules where nuclear motion is slow. Therefore in our calculation, the molecular internuclear distance is frozen. We present two sets of Molecular QPI results with this calculation. In one of them only a strong 800 nm laser field, polarized perpendicularly to the molecular axis, is present. In the other one an additional resonant field, with $3\omega$ carrier frequency, polarized parallel to the molecular axis, is added. In theoretical analysis the harmonic emission associated to different channels, ionisation from e.g. HOMO and HOMO-1, can be resolved \cite{suren_pra}.

In Fig. 7 we show as an example the results obtained for the 15th harmonics. QPI pattern for separated HOMO (ground state) and HOMO-1 (excited state) with and without the resonant field that couples these two states. In the case of the absence of the resonant field, the HOMO (blue dashed line) and HOMO-1 (green solid line) QPI patterns are not shifted with respect to each other, since the HOMO-1 state of the ion is populated via the excitation by the recolliding electron only \cite{suren_prl}, which gives no additional accumulated phase. While adding the resonant field the QPI pattern calculated for HOMO-1 (solid red line) presents a shift by quasi $\pi$ with respect to the HOMO (black dashed line) due to the additional phase caused by the resonant population transfer between the cation states. The resonant transition is tuned to exceed the transitions via the recollision, by choosing an appropriate intensity for the coupling field. Therefore the contrast of the HOMO-1 and HOMO populations in the presence of resonant field will affect strongly the contrast of the full Molecular QPI signal and is a signature of the population transfer.

\begin{figure}
\centering
\includegraphics [width=3.5 in,height=3 in]{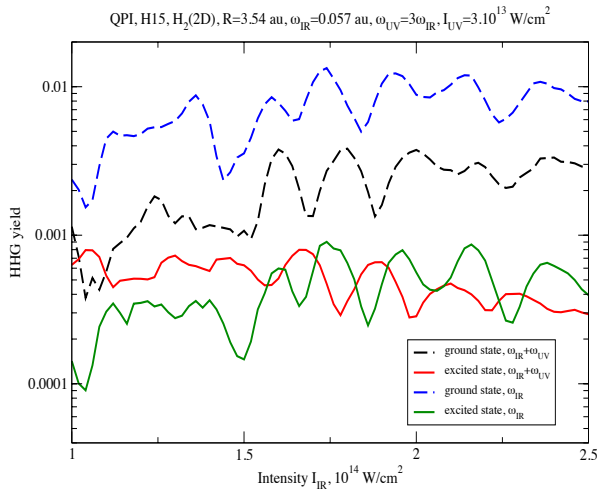}
\caption{QPI pattern for 15th harmonic for a model molecule of fixed internuclear distance, R = 3.54 a.u . 
The lines are associated to HOMO without ( blue dashed line) and with (black dashed line)
the resonant field, and to HOMO-1 without (green solid line) and with ( red dashed line) the resonant field.
The strong laser field is polarized perpendicular and the resonant laser field parallel to the 
molecular axis.}
\label{Fig.7}
\end{figure}

\section{Conclusion}
In conclusion, we have performed studies that demonstrate the potential of Quantum-path Interferences (QPI) to probe complex dynamics now in molecules. Our first study is experimental and compared short and long trajectories in randomly aligned molecules ($CO_{2}$, $N_{2}$, $O_{2}$, $D_{2}$, $H_{2}$). We have clear evidence that the ratio of short and long trajectories is sensitive to nuclear motion of the cation with attosecond-femtosecond time scale, especially in $H_{2}$ and $D_{2}$ cases where the comparison is direct. We also show that QPI, in randomly aligned molecules is governed by the continuum electron phase (as in the atomic case) and we highlight the fact that the electronic dynamic in the continuum during HHG process is independent on the non-isotropic molecular potential (allowing the use of strong field approximation for Molecular QPI). In a second theoretical study we extend the QPI to aligned molecules in order to get inside the electronic cation dynamics. We study the cases of multiple channels of ionisation contribution as well as population transfer due to laser field coupling. In a first simple approach based on atomic SFA, we show that the multiple orbitals contribution to the Molecular QPI should appear as a shift of $\pi/2$  of the modulation towards low intensities. In a quasi-analytical approach based on nitrogen SFA including two ionic states symmetries and their coupling by the laser field in perpendicular alignment, the Molecular QPI pattern is clearly shifted towards the lower intensity as for the atomic model but also shows QPI contrast losses.This indicates that the shift is a clear signature of the multiple channels contribution and the lack of contrast shall allowed to identify the population transfer between the ionisation states HOMO and HOMO-1. In a third step we calculate the Molecular QPI in a model $H_{2}$molecule where the population transfer between ionic states is allowed by a resonant field. In that case the QPI pattern is calculated for each individual channels. The HOMO-1 channel reveals a clear shift by quasi $\pi$ due to the additional phase caused by the population transfer between the cation states. It will therefore affect strongly the contrast of the full QPI as suggested by the quasi-analytic model.

Hence the QPI method appears to be an efficient method to reveal the details of the intra-molecular dynamics in a self-referenced way and applicable to many type of molecules.

\begin{acknowledgments}
This work is supported by the UK EPSRC  (Grant No. $EP/J002348/1$, $EP/I032517/1$  and $EP/E028063/1)$ , $ERC-2011-ADG_20110209 Project 290467 (ASTEX)$, and the British Council Alliance 10026. The authors would like to thank also Andrew Gregory and Peter Ruthven for technical support.
A.Z would like to thanks Davide Fabris and Felicity McGrath, for following the experiments and the calculations.
A.Z would like to thanks Roland Guichard, Jeremie Caillat, Richard Taieb and Alfred Maquet for fruitful discussion on plane wave approximation and molecular alignment.
\end{acknowledgments}









%




%











%












\end{document}